Sampling Issues in Bibliometric Analysis

Richard Williams** and Lutz Bornmann*

Last revised November 16, 2015

Corresponding author:

**Department of Sociology

810 Flanner Hall

University of Notre Dame

Notre Dame, IN 46556 USA

E-mail: Richard.A.Williams.5@ND.Edu

Web Page: http://www3.nd.edu/~rwilliam/

*Division for Science and Innovation Studies

Administrative Headquarters of the Max Planck Society

Hofgartenstr. 8,

80539 Munich, Germany.

E-mail: bornmann@gv.mpg.de

**Abstract**

Bibliometricians face several issues when drawing and analyzing samples of citation records for their research. Drawing samples that are too small may make it difficult or impossible for studies to achieve their goals, while drawing samples that are too large may drain resources that could be better used for other purposes. This paper considers three common situations and offers advice for dealing with each. First, an entire population of records is available for an institution. We argue that, even though all records have been collected, the use of inferential statistics, significance testing, and confidence intervals is both common and desirable. Second, because of limited resources or other factors, a sample of records needs to be drawn. We demonstrate how power analyses can be used to determine in advance how large the sample needs to be to achieve the study's goals. Third, the sample size may already be determined, either because the data have already been collected or because resources are limited. We show how power analyses can again be used to determine how large effects need to be in order to find effects that are statistically significant. Such information can then help bibliometricians to develop reasonable expectations as to what their analysis can accomplish. While we focus on issues of interest to bibliometricians, our recommendations and procedures can easily be adapted for other fields of study.

# 1    Introduction

Statistical significance tests and/or confidence intervals (CIs) are frequently used with bibliometric data. For example, Opthof and Leydesdorff (2010) compared leading scientists (professors) at the Academic Medical Center of the University of Amsterdam using the Kruskal-Wallis test. Statistical significance tests are strongly connected to questions of sampling, since these tests are usually applied to the analysis of samples in order to obtain information about an underlying population (Levy & Lemeshow, 2008). In bibliometrics, several papers have been published which deal with the use of significance tests and effect sizes (e.g. Bornmann & Williams, 2013; Schneider, 2012; Schneider, 2013), but the literature on sampling of populations is scarce. In one of the rare papers, Bornmann and Mutz (2013) argue for clusters in a two-stage sampling design ("cluster sampling"), in which, firstly, one single cluster is randomly selected from a set of clusters (e.g. consecutive publication years, in which an institution have published) and secondly, all the bibliometric data (publications and corresponding citation metrics) is gathered (census) for the selected cluster. Then, this cluster sample can be statistically analyzed.

This paper deals with issues around samples and populations in bibliometrics. In many institutional evaluations, bibliometricians have complete publication and citation records for all the papers of an institution. These are sometimes referred to as "apparent populations;" Berk, Western, and Weiss (1995b) give as examples of apparent populations all states in the United States or all nations in the developing world. We argue that, even though all records have been collected, the use of inferential statistics and significance testing is both common and desirable. We further argue that the use of power analysis can help guide analyses when records for an entire population are not available. Specifically, this paper addresses two issues: first, the appropriateness of using inferential statistics when the entire population of records is available (Bornmann, 2013); and second, the use of power analysis and sampling when it is impractical to gather information for all institutional citation records (Bornmann & Mutz, 2013). In particular, how does a bibliometrician go about determining how large a sample needs to be in order to achieve the goals of the analysis? Conversely, when the sample size has already been determined, how large do effects need to be in order for them to be statistically significant? Answering such questions can help the bibliometrician decide how large a sample is needed; or, if the sample has already been drawn, answering these questions can help the bibliometrician form reasonable expectations as to what the analysis can accomplish.

# 2 Justification for using Statistical Inference with Citation Impact Data

## 2.1 Appropriate types of data

In the following, we discuss techniques that are appropriate when a study wishes to use percentiles of citations to measure institutional citation impact. We also note that, while we focus on the analysis of percentile data, our ideas could also be applied to other types of bibliometric statistics, such as statistics based on average citations rather than percentile-based statistics.

Cross-field and cross-time-period comparisons of citation impact for institutional evaluation purposes are only possible if the impact is normalized (standardized) (Schubert & Braun, 1986). For its citation impact to be normalized, a paper needs to have a reference set: all the papers published in the same publication year and subject category. Percentiles have been proposed as a robust alternative to normalization on the basis of central tendency statistics (arithmetic averages of citation counts) (Hicks, Wouters, Waltman, de Rijcke, & Rafols, 2015; Wilsdon et al., 2015). Percentiles are based on an ordered set of publications in a reference set, whereby the fraction of papers at or below the citation counts of a paper in question is used as a standardized value for the relative citation impact of this focal paper. This value can be used for cross-field and cross-time-period comparisons. If the normalized citation impact for more than one paper is needed in a research evaluation study (and this is the rule in institutional evaluations), this percentile calculation is repeated (by using corresponding reference sets for each one).

Following the practice of Incites (Thomson Reuters, http://incites.thomsonreuters.com/), we use inverted percentiles in our examples, where low percentile values mean high citation impact. Hence citation impact above the median (in the field and publication year) is defined as percentiles less than 50. With inverted percentiles it can easily be seen whether a citation percentile is within the top 10 or top 1 percent most frequently cited paper range, which we think is the sort of thing most bibliometricians will be interested in (Bornmann, 2014). Of course it is a trivial matter to use non-inverted percentiles instead if the bibliometrician prefers them or if it is appropriate given the way the data being analyzed are coded.

## 2.2    Using bootstrapping to verify that the statistical methods employed are appropriate for percentile data

A possible statistical problem in this study is that percentiles have an approximately uniform rather than normal distribution.[1] When variables are normally distributed, cases tend to be clustered near the mean, while extreme values in either direction are less common. With percentile rankings, however, in the population there will be just as many cases in the first percentile as there are in the 50th and the 99th. t tests assume that dependent variables are normally distributed, which raises the question of whether analyses based on t tests (which includes the power analyses presented here) are potentially biased.

---

[1] Their distribution is uniform only approximately, depending on the number of ties in the citation distribution.

A recent analysis by Williams and Bornmann (2014) suggests that a power analysis of percentile rankings can indeed be conducted. Bootstrapping is often used as an alternative to inference based on parametric assumptions when those assumptions are in doubt (Cameron & Trivedi, 2010). Bootstrapping resamples observations (with replacement) multiple times. Standard errors, CIs and significance tests can then be estimated from the multiple resamples. Using real data for the years 2001 and 2002 from three research institutions in German-speaking countries, Williams and Bornmann (2014) made heavy use of t-tests and related statistics. They used bootstrapping to double-check their results, and found that "bootstrapping produced significance tests and confidence intervals that were virtually identical to those reported in our tables, giving us confidence that our procedures are valid" (p. 269). We therefore feel confident that the statistical techniques we use in this paper are appropriate and that our findings are valid.

## 2.3    Statistical Software

For the calculation of the statistical procedures in this paper, we used Stata (StataCorp, 2013)[2]. Appendix A contains the codes used. However, many other statistical software routines could also be used for these calculations (e.g. SAS or R).

# 3    The Use of Power Analysis under Different Sampling Conditions

We consider three common situations. First, an entire population of records is available for an institution. Second, because of limited resources or other factors, a sample of records needs to be drawn. Third, the sample size may already be determined, either because the data have already been collected or because resources are limited.

---

[2] In particular, we used the power and sample size routines included with Stata 13. These include such programs as onemeans and twomeans as well as several other types of routines for methods not used in this paper.

## 3.1    Using inferential statistics to analyse a population

It could be argued that there is no need to compute significance tests or CIs given bibliometric population data for an institution. That is, we do not need to estimate parameters or make inferences about the larger population because the information on the entire population of papers is available. For example, do we really need to use CIs to estimate a range of plausible values for the mean when we already have all the information to determine what the population mean is? By way of analogy, a public opinion poll may estimate, subject to some degree of sampling error, who is leading in an election. But, once the election has been held we no longer need to estimate the levels of support because we know who actually got the most votes.

However, in situations similar to institutional evaluations, it is actually quite common to go ahead and perform significance tests and compute CIs anyway. Bielby (2013), for example, notes that significance tests are widely used in class action employment lawsuits even when all employee records are available for analysis. Two rationales are typically offered for treating what appears to be a population as though it were a sample.

First, the current cases might be thought of as being a sample from a larger population that includes future cases as well (Gelman, 2009). As Bollen (1995) notes, data from one year may be followed by data from later years. Hence, you really do not have the entire population, you just have information from one or more points in time. This is not a new argument; Deming and Stephan (1941) made a very similar claim back in 1941. They said (p. 45) that

As a basis for scientific generalizations and decisions for action, a census is only a sample. In addition to serving the function of an inventory as of a certain date, the census tabulations serve also another important objective, namely, as bases for prediction. Any social or economic generalization, and any recommendation for a course of action, involves a prediction. For such purposes, the census takes on the character of a sample.

Researchers from Canada's Manitoba Center for Health Policy (2001, p. 1) reached similar conclusions:

> [A majority of us] reached the conclusion that even when one has data on the full population, one only has that data cross-sectionally in time. In a sense, the data can be viewed as a sample from possible states in the Province as they unfold over time. Therefore, it made sense to us to try to indicate whether differences which are certainly real across units are statistically significant when one considers the data to be a one-time sample of the unfolding of the universe.

A second rationale, and a perhaps more compelling one, is to think of observed cases as repeated trials that are products of an underlying stochastic process. If we tossed a coin 100 times, we would not think that we had the entire population of coin tosses; a different set of tosses is possible and, because of chance factors, would likely yield somewhat different results. As Berk et al. (1995b, p. 423) explain,

> …the data are treated as a 'realization' of some set of social process that could have in principle produced a very large number of other realizations. These realizations, in turn, constitute a super population. That is, the data could have been different as a result of random sampling from the 'super population.' Then, conventional statistical inference is applied as usual. An apparent population has now become a random sample.

Again, Deming and Stephan (1941, pp. 45-46) made basically the same argument more than 70 years ago:

> The births, deaths, vocations, migrations, and educational attainments of a population are changed and directed by a myriad of chance causes, superimposed on certain underlying social and economic cause systems. A census shows what resulted from this combination at a certain time in the past, but any generalizations that are not restricted to a particular date and place must recognize the fact that some other population might have resulted, and must in fact be expected to arise in the future from the same underlying causes. Because of these statistical fluctuations, it follows that as a basis for scientific generalizations and decisions for action, the distinction between complete and sample coverages is often only a matter of degree.

Bollen (1995) extends this argument. He says that the dependent variable in an analysis, even with complete population data, cannot be predicted perfectly. Models therefore include unmeasured random variables, which may reflect all the unmeasured influences on the dependent variable. Further there may be measurement errors in the data which also adds stochastic elements to the analysis. Bollen (1995, p. 467) therefore argues that

Another strategy is to employ the super-population model. In the regression context this means that we view the disturbance term as a random variable for which we have a finite number of realizations. And from that finite number, we attempt to make statistical inferences about the generating distribution.

So, for example, in an employment discrimination case, if women are making less than men it might be that chance factors caused some women to be unlucky with their wages (like tossing a coin and getting five tails in a row) even though overall the process by which wages are set is fair.

For bibliometrics, we argue that the observed citation impact of papers (measured by percentiles) allows us to make inferences about the underlying process that generated those impacts and the extent to which citations may have been influenced by random factors. The success of a paper, or of an entire institution, is presumably affected by the quality of the material in the papers, but is also partly determined by chance. For example, how often a paper or collection of papers gets cited might be affected by how many people chose to read a particular issue of a journal or who happened to learn about a paper because somebody casually mentioned it to them (Cronin, 1984). Put another way, if we could somehow repeat the citation process over and over (e.g. by publishing the paper in different issues of the same journal), the citation impact of papers (percentiles of citations) would not be exactly the same for each repetition, just like doing 100 coin tosses over and over would not yield the exact same number of heads each time. Hence, even when all existing citation records for an institution are available, inferential methods can still be used to test whether, say, a high impact score for an institution could just be due to luck, or whether apparent differences in the average percentiles for two institutions are too large to attribute to chance alone.

Before closing this section, we should note that not everyone agrees with the arguments presented her. For example, Schneider (2012, p. 728) contends that super-populations are "very often assumed, but seldom justified". Schneider (2015, p. 411) further argues that "many researchers are not aware of the numerous criticisms raised against NHST" (null hypothesis statistical significance tests). Similarly, Berk, Western, and Weiss (1995a, p. 452) say that "statistical inference is very commonly used when it obscures far more than it enlightens." Behnke (2005) also urges caution when using methods of statistical inference to study entire populations and offers criteria for when it may be appropriate. For those interested in hearing more, *Sociological Methodology* 1995 contains a lively back and forth discussion on this topic (e.g. Berk et al., 1995a, 1995b; Bollen, 1995; Firebaugh, 1995; Rubin, 1995).

## 3.2    Using power analysis when a sample needs to be drawn

While it is desirable to have all institutional records for an evaluation study it is not always practical. Percentile data need to be purchased from databases (e.g. from Scopus provided by Elsevier), and the amount of money and/or effort required for obtaining percentiles for all records may be prohibitive. For example, Scopus provides percentiles for every single publication, and most scientists have access to this database. However the search for percentiles is time-consuming. With Scopus one cannot download the percentiles for a larger publication set. Instead one has to search and display every single publication from a data set (e.g. publications of an institution) and has to write down the percentile. Thus, unless a great deal of inexpensive labor is available, if the user decides to use Scopus a sample should be as small as possible. InCites (Thomson Reuters) and SciVal (Elsevier) offer percentiles for larger sets, but these are expensive products designed for institutional bibliometric analyses.

In other cases, a bibliometrician may wish to supplement the (citation impact) information contained in the bibliographic records; for example, add information about the authors (e.g. their academic status) or more refined codings of the topic matter. It may be impractical or too expensive to do this for all the records and hence a sample will need to be selected.

However, how does a bibliometrician decide how big a sample needs to be drawn? Samples that are either too small or unnecessarily large both have disadvantages. As StataCorp (2013) notes (see also Cohen, 1988), "a study with too few subjects may have a low chance of detecting an important effect, and a study with too many subjects may offer very little gain and will thus waste time and resources" (p. 1). To determine optimal sample size, power analyses are often conducted before a sample is collected. A typical use of power analysis is to determine how large the sample must be to detect an effect of a given size. That is, how large does the sample need to be that we can be reasonably confident that we will correctly reject the null hypothesis when the null hypothesis is false?

So, for example, suppose that an institution believes that it is above average in terms of how often its publications get cited. If the papers of the institution really are above average, how much above average does it need to be, and how large does the sample need to be, in order to detect statistically significant differences from the average score in the reference sets (percentile=50)? A power analysis can be used to address such questions (see Table 1).

A technical explanation of the mathematics behind power analysis is beyond the scope of this paper[3], but we can explain several of the key components behind such an analysis. We interpret the results as follows:

- 28.87 is the approximate population standard deviation ($\sigma$) for percentile rankings (Waner & Costenoble, 1996)[4]. It is common to assume that the sample standard deviation will be the same, although the bibliometrician could choose some other value if there were reason to believe otherwise. The value of 28.87 is used in all of the analyses presented here.

- Power ($\pi$) = P(rejecting h0| h0 is false). For example, if we set power at .8, that means that we want a sample size that is large enough that we will correctly reject the null 80% of the time when it is false. If more power is deemed necessary a larger value can be chosen, but this will also require a larger sample size.

- Alpha ($\alpha$) = P(rejecting h0| h0 is true). $\alpha$ = .05 is a commonly used criterion for rejection; differences between the null and alternative hypotheses must be large enough that we would expect to only reject the null 5% of the time when the null is true. More stringent (e.g. .01) or less stringent (.10) values can be chosen, depending on how costly we feel it would be to reject the null when we should not.

---

[3] Numerous other sources, such as Cohen (1988) can be consulted by those wishing to see a more technical and mathematical discussion.

[4] More specifically, percentiles from InCites have an approximately uniform distribution with values ranging from 0 to 100. As Waner and Costenoble (1996) and others note, the formula for the standard deviation of a variable with a continuous uniform distribution is (highest value − lowest value) / $\sqrt{12}$ = 100/ 3.464 = 28.87. Since percentile rankings technically have a discrete distribution the value of 28.87 can be viewed as an approximation (but probably a very good one).

- $\mu_0$ is the value of the mean specified under the null hypothesis. For all of our analyses we chose the known population mean of 50, but we could have chosen higher or lower values if we had felt they were more appropriate. For example, a major research institution that considers itself among the world's elite might want to see whether it exceeds a more demanding value like 25. Conversely, a teaching oriented regional college might feel that a more modest value like 75 is appropriate.

- $\mu_a$ is the hypothesized alternative value for the mean. In this case we specify differences from the mean that are as small as 2.5% and as large as 10%. The smaller the hypothesized difference, the larger the sample size needs to be in order to be reasonably confident that a false null hypothesis will be rejected.

- Delta ($\Delta$) is a standardized measure of effect size, which equals $(\mu_a - \mu_0)/\sigma$. So, for example, when $\mu_a = 47.5$, $\Delta = (47.5 - 50)/28.87 = -.0866$. The larger the effect size is, the smaller the sample needs to be to produce statistically significant effects. This and other standardized measures can be useful when it is not otherwise clear how substantively significant differences are. If, for example, we knew that students in an experimental teaching program scored one grade level higher than their counterparts in traditional programs, such a difference might have a great deal of intuitive meaning to us. But if instead we knew that they scored 7 points higher on some standardized test, effect size measures could help us to assess how important such a difference really is. Bibliometricians may have a clear idea of whether or not being 5 points above average is substantively important, but if not measures of effect size can help to guide the analysis and the sample selection.

- N is the sample size that is needed, given the values that have been specified for $\alpha$, $\pi$, $\mu_0$ and $\mu_a$, and $\sigma$. In this case N is estimated while the other values have been specified by the researcher. As shown later it is possible to instead fix the value of N (e.g. set the sample size at 200) and then estimate other quantities, e.g. how much power does the sample have?

The results in Table 1 tell us that the sample size needs to be 1,049 or greater to be reasonably confident that a real difference from 50 (the population average) of as little as 2.5 points will be found to be statistically significant. A 5 point difference only requires a sample size of 264, and a difference as large as 10 points only requires a sample size of 68. Hence, a bibliometrician who felt that only differences of five points or greater were worth caring about might choose to draw a much smaller sample than a bibliometrician who felt that a difference of as little as 2.5 points was important.

If the institution wants to collect a smaller sample, it could specify a higher value for α (e.g. .10) or a lower value for the power. A smaller sample will increase the chances of rejecting the null when we should not or accepting the null when it is false. Conversely, if we had the resources and wanted more precise and powerful estimates, we could make α smaller (e.g. .01) and/or make power higher (e.g. .9). As the results in Table 2 show, to meet both of these more stringent standards sample sizes would have to be almost twice as large as before.

### 3.3 Using power analysis when a sample has already been drawn: target means and minimum detectable differences

There may also be situations in which the sample size is already known. Perhaps the data have already been collected; or, available resources only allow the collection of a limited number of records. In such instances, bibliometricians may wish to know what the smallest possible effect and corresponding "target mean" (i.e. the mean of the sample) will have to be in order to detect statistically significant results when the null is false. This is also referred to as the "minimum detectable difference." So, for example, suppose an institution can only afford to collect percentile data for 200 cases in Scopus. Table 3 shows how large differences from the average score in the reference sets will have to be in order to achieve statistical significance.

The results in Table 3 show that, in order for the bibliometrician to be reasonably confident that results would be statistically significant at the .01 level, the true mean for the institution would need to be more than 7 points better (42.96) than the average percentile in the reference set (50). Using the .05 level, the institution would still need to average almost 6 points better (44.25). The less demanding .1 level of significance would require a real difference of slightly over 5 points (44.91). If the institution correctly believed that it was 4 points better than average, a sample size of only 200 would not be large enough to reasonably guarantee that the institution's mean would be found to be statistically significantly better than the average impact in the reference sets. If this is not considered acceptable the bibliometrician may wish to choose a less demanding value for $\alpha$ or, better yet, see if there is some way for additional data to be collected.

The above sorts of calculations can also be useful even when records for all publications have been collected. For example, an institution with relatively few publications can determine how much above average it has to be in order to expect statistically significant results. A power analysis may show that, even if an institution is above average, a statistical analysis is unlikely to yield statistically significant results. Conversely, for a larger institution, a power analysis may reveal that even trivial differences from the average in the reference sets are likely to be statistically significant.

## 3.4 Other possible analyses

Similar calculations can be done for other purposes. We might want to know how large sample sizes need to be to detect differences between two institutions, or how large the sample needs to be to see whether an institution has an exceptionally large number of "excellent" publications, e.g. publications that rank among the top 10% in the corresponding subject category and publication year.

At the same time several factors can make power analyses more complicated. Several analyses involving different variables may be planned, and the optimal sample sizes for each may differ. If subsample analyses are also planned (e.g. papers in certain fields only) that too needs to be taken into account when determining sample size, i.e. sample sizes for each subsample must also be large enough to achieve the study's goals. Assumptions made in the calculations (e.g. sample standard deviations) may prove to be inaccurate, causing the original calculations of needed sample sizes to be too optimistic or pessimistic. In order to ensure that sample sizes are sufficiently large bibliometricians may wish to choose somewhat more stringent values for power and α.

Finally, while we have focused on issues of interest to bibliometricians, similar concerns about samples and sample size occur in many areas of research. Our recommendations and procedures can easily be adapted for other fields of study.

## 4    Discussion

Bibliometricians will sometimes enjoy the luxury of having complete records for an institution. However, even in such cases the use of inferential statistics is appropriate and helpful. The observed values did not have to come out as they did. Chance factors could have increased the number of citations a paper received or else decreased them. Further, even when all records are available, a power analysis can be useful for determining what the reasonable expectations are for the study. A power analysis can indicate how difficult it is to get statistically significant results even when the citation impact of a small institution's publications really is above average (or 50% above average, or whatever criterion for evaluation the institution has set for itself) in the corresponding reference sets; or conversely, how easy it is for the citation impact of a large institution's publications to achieve statistically significant results even if the substantive differences between their citation impact and the impact of the publications in the corresponding reference sets are trivial.

In other situations, a sample will need to be drawn. Before drawing the sample, it is important to assess how large the sample needs to be to achieve its goals and provide the best allocation of resources. If an institution feels that it is about 4 points above average, then even if it is right a sample that is too small may fail to support its beliefs. There is little point in conducting a study if it is likely doomed to failure before it even gets started. But, if an institution spends money collecting far more data than is necessary, it may have to cut back on expenditures in other important areas, e.g. data analysis. The examples and guidelines provided in this paper can help guide bibliometricians when deciding how large their samples ought to be and what they can reasonable expect from their data once they have it.

Engagement in inferential statistics and power analysis results in thinking about meaningful bibliometric results. The results of citation analyses (e.g. in university rankings) are often presented in a precise form and minimal differences are used to rank institutions. The presentation of ranking positions (calculated from bibliometric results) dispenses the user from the examination of the initial results which lead to the rankings. However, this examination is encouraged with the use of inferential statistics and power analyses: The user has to think about the meaning of a particular difference between the citation impact scores of two institutions.

**Appendix A: Stata 13.1 code**

```
* Table 1
power onemean 50 (47.5(-2.5)40), sd(28.87) table graph(name(Fig1, replace))

* Table 2
power onemean 50 (47.5(-2.5)40), sd(28.87) alpha(.01) power(.9) ///
      graph(name(Fig2, replace)) table

* Table 3
power onemean 50, sd(28.87) n(200) power(.8) direction(lower) ///
      alpha (.01 .05 .1) graph(name(Fig3, replace)) table
```

Table 1. Estimated sample size for a one-sample t test*

| Alpha (α) | Power (π) | N | Delta (Δ) | $\mu_A$ |
|---:|---:|---:|---:|---:|
| .05 | .8 | 1049 | -.0866 | 47.5 |
| .05 | .8 | 264 | -.1732 | 45 |
| .05 | .8 | 119 | -.2598 | 42.5 |
| .05 | ,8 | 68 | -.3464 | 40 |

* $\mu_0 = 50$ and $\mu_0 = 28.87$ in all analyses.

Table 2. Estimated sample size for a more stringent one-sample t test*

| Alpha ($\alpha$) | Power ($\pi$) | N | Delta ($\Delta$) | $\mu_A$ |
|---:|---:|---:|---:|---:|
| .01 | .9 | 1988 | -.0866 | 47.5 |
| .01 | .9 | 500 | -.1732 | 45 |
| .01 | .9 | 224 | -.2598 | 42.5 |
| .01 | ,9 | 128 | -.3464 | 40 |

* $\mu_0 = 50$ and $\mu_0 = 28.87$ in all analyses.

Table 3. Estimated target mean for a one-sample mean test*

| Alpha ($\alpha$) | Power ($\pi$) | N | Delta ($\Delta$) | $\mu_A$ |
|---|---|---|---|---|
| .01 | .8 | 200 | -.2437 | 42.96 |
| .05 | .8 | 200 | -.1991 | 44.25 |
| .10 | .8 | 200 | -.1764 | 44.91 |

* $\mu_0 = 50$ and $\mu_0 = 28.87$ in all analyses.